\documentclass[12pt,preprint,tightenlines,prd]{revtex4}

\begin{document}
\begin{flushright}
 \textbf{CERN-TH/2002-178} \\
\end{flushright}

\title{\vspace*{24pt}
\LARGE\bf Yang-Mills Theory for Non-semisimple Groups\\[24pt]}

\author{\large\bf Jean \surname{Nuyts}}
\thanks{E-mail: Jean.Nuyts@umh.ac.be}
\affiliation{Universit\'e de Mons-Hainaut,
20 Place du Parc, 7000 Mons, Belgium\vadjust{\kern10pt}}

\author{\large\bf Tai Tsun \surname{Wu}}
\thanks{E-mail: ttwu@deas.harvard.edu
\vfil
\vskip.5in
\begin{flushleft}
 \normalsize \textbf{CERN-TH/2002-178}\\
 \textbf{September 2002}
\end{flushleft}
}
\affiliation{Gordon McKay Laboratory, Harvard University,
Cambridge, MA 02138, U.S.A., and
Theoretical Physics Division, CERN, CH-1211 Geneva 23,
Switzerland
\vspace{30pt}}

\begin{abstract}
For semisimple groups, possibly multiplied by U(1)'s, the number of
Yang-Mills gauge fields is equal to the number of generators of the group.
In this paper, it is shown that, for non-semisimple groups, the number of
Yang-Mills fields can be larger. These additional Yang-Mills fields are not
irrelevant because they appear in the gauge transformations of the original
Yang-Mills fields.  Such non-semisimple Yang-Mills theories may lead to
physical consequences worth studying. The non-semisimple group with only
two generators that do not commute is studied in detail.
\end{abstract}
\maketitle 
\thispagestyle{empty}
\eject

\section{Introduction}\label{sec:1}

The non-Abelian gauge field theory was invented by Yang and Mills \cite{YM}
almost half a century ago; it now permeates the study of elementary
particles, both strong and electroweak. In all cases, the number of spin-1
gauge fields is equal to the number of generators of the gauge group. For
example, for the SU(2) case, there are three generators and also three
Yang-Mills fields. It is the purpose of the present paper to study the
question: What happens if more Yang-Mills fields are introduced than the
number of generators of the gauge group?

This possibility has actually been investigated already in the original
paper of Yang and Mills \cite{YM}. They showed that such additional gauge
fields are ``allowed by the very general form'' but ``irrelevant to the
question of isotopic gauge.''

It is this sentence of Yang and Mills that initiated the present
investigation. Since their conclusion holds whenever the gauge group is
semisimple, the additional gauge fields are of interest for non-semisimple
gauge groups. It is found that, at least for some cases and perhaps in most
cases, the number of relevant spin-1 gauge fields can be larger than the
number of generators of the underlying gauge group. It is hoped that this
phenomenon, which we did not anticipate, may be of use for particle physics.

In particle physics, the ``simplest'' Lie groups seem to play the most
fundamental roles. For example, in the electroweak theory of Glashow
\cite{G}, Weinberg \cite{W} and Salam \cite{S}, the group is
SU(2)$\otimes$U(1). It is therefore the purpose here to study the
``simplest'' non-semisimple group (besides U(1)$^n$) in the sense that the
number of generators is the smallest. If there is only one generator, then
the Lie group is necessarily Abelian. We shall concentrate here on the
non-Abelian Lie group with two generators.

These two generators $L_1$ and $L_2$ can be chosen to obey the commutation
relation
\begin{equation}
\left[L_1,L_2\right]= L_2\ .
\label{algebra}
\end{equation}
This commutation relation leads essentially to only one gauge group.
Faithful representations of the lowest dimension are
\begin{eqnarray}
L_1&=&\left(\matrix{\frac{1}{2}+s & 0          \cr 
                    0       &-\frac{1}{2}+s     }
    \right)\ ,
    \nonumber\\
L_2&=&\left(\matrix{0       & 1     \cr 
                    0       & 0     }
    \right)\ ,
\label{rep2}
\end{eqnarray}
where $s$ is an arbitrary complex parameter.

In Sec.\ \ref{sec:2}, we study the transformation properties of the gauge
fields relevant for a doublet of scalar fields transforming locally with the
representation (\ref{rep2}). In Sec.\ \ref{sec:3}, we discuss the
elementary properties of these gauge fields and especially the influence of
the $s$ parameter. In Sec.\ \ref{sec:4}, we extend our results to all
representations where $L_1$ can be diagonalized and show in Sec.\
\ref{sec:5} how the gauge system of the 2-dimensional representation
extends directly to all these representations. In Sec.\ \ref{sec:6}, the
Lagrangian of the gauge fields is constructed on general grounds. Finally,
we give a brief discussion and conclusions in Sec.\ \ref{sec:7}.

\section{Gauge Transformations}\label{sec:2}

Following step-by-step the procedure pioneered by Yang and Mills, we
consider a doublet scalar field in four space-time dimensions
\begin{equation}
\Phi(x)=\left({\matrix{\Phi_1(x)\cr \noalign{\vskip3pt}
                      \Phi_2(x)}}\right)
\label{doublet}
\end{equation}
that transforms locally as
\begin{equation}
\Phi^{\prime}(x)=V(x)\Phi(x)\ .
\label{Phitf}
\end{equation}
The infinitesimal form of $V(x)$ is
\begin{equation}
V(x)=1+\alpha_1(x) L_1+\alpha_2(x) L_2\ ,
\label{Vtf}
\end{equation}
where $L_1$ and $L_2$ are given by Eq.\ (\ref{rep2}).

The derivative $D_{\mu}\Phi$ is defined by
\begin{equation}
D_{\mu}\Phi=\left(\partial_{\mu} +A_{\mu}\right)\Phi
           =\partial_{\mu}\Phi +A_{\mu}\Phi
\label{covder}
\end{equation}
and must transform in the same way as $\Phi$ itself
\begin{equation}
\left(D_{\mu}\Phi\right)^{\prime}(x)
=V(x)\,(D_{\mu}\Phi)(x)\ .
\label{covdertf}
\end{equation}
Equations (\ref{Phitf}) and (\ref{covdertf}) imply that
\begin{equation}
A^{\prime}_{\mu}=VA_{\mu}V^{-1}-(\partial_{\mu}V)V^{-1}\ ,
\label{Atf}
\end{equation}
meaning that this basic result of Yang and Mills is valid for the present
gauge group.
The infinitesimal form of Eq.\ (\ref{Atf}) is
\begin{equation}
A^{\prime}_{\mu}=A_{\mu}
                +\alpha_1\left[L_1,A_{\mu}\right]
                +\alpha_2\left[L_2,A_{\mu}\right]
                -\partial_{\mu}\alpha_1 L_1
                -\partial_{\mu}\alpha_2 L_2
      \ .
\label{Atfinf}
\end{equation}

Let us study this equation in some detail. Since $\alpha_1$ and $\alpha_2$
are arbitrary functions of the space-time variables $x$, this equation
implies that there must be at least two gauge fields in $A_{\mu}$. The
usual choice is
\begin{equation}
A_{\mu}(x)=A_{\mu}^{(1)}(x) L_1+A_{\mu}^{(2)}(x) L_2\ .
\label{usual}
\end{equation}
With this choice, which involves two gauge fields $A_{\mu}^{(1)}(x)$ and
$A_{\mu}^{(2)}(x)$, Eq.\ (\ref{Atfinf}) can indeed be satisfied.

Following the discussion of Yang and Mills as quoted in Sec.\ \ref{sec:1},
Eq.\ (\ref{usual}) is not the only possible choice: it is entirely allowed
to have more than two gauge fields. As seen from Eq.\ (\ref{covder}) or
Eq.~(\ref{usual}) for example, $A_{\mu}(x)$ is a $2\times 2$ real matrix,
and it is therefore natural to consider the case of {\textit{four}} gauge 
fields, namely $A_{\mu}^{ij},\ i,j=1,2$. It is convenient to organize these
four $A_{\mu}^{ij}$ as a column matrix $A_{\mu}^{a},\ a=1,2,3,4$:
\begin{equation}
\left(\matrix{A^1_{\mu}\cr \noalign{\vskip3pt}
              A^2_{\mu}\cr \noalign{\vskip3pt}
              A^3_{\mu}\cr \noalign{\vskip3pt}
              A^4_{\mu} } \right)
\equiv
\left(\matrix{A^{11}_{\mu}\cr \noalign{\vskip3pt}
              A^{12}_{\mu}\cr \noalign{\vskip3pt}
              A^{21}_{\mu}\cr \noalign{\vskip3pt}
              A^{22}_{\mu} }\right)\ .
\label{reading}
\end{equation}
In this notation, the transformations of Eq.\ (\ref{Atfinf}) are
\begin{equation}
{A}^{\prime}_{\mu}={A}_{\mu}
+\alpha_i {X}_i {A}_{\mu}
+\partial_{\mu}\alpha_i {W_i}
\label{Afourtf}
\end{equation}
with
\begin{eqnarray}
   {X}_1&=&\left(\matrix{ 0        & 0         & 0      & 0          \cr
                          0        & 1         & 0      & 0          \cr
                          0        & 0         & -1     & 0          \cr
                          0        & 0         & 0      & 0       }
         \right)\ ,
      \label{xonefour}\\[8pt]
   {X}_2&=&\left(\matrix{ 0        & 0         & 1      & 0          \cr
                          -1       & 0         & 0      & 1          \cr
                          0        & 0         & 0      & 0          \cr
                          0        & 0         & -1      & 0     }
         \right)\ ,
      \label{xtwofour}\\[8pt]
   {W_1}&=&\left(\matrix{  -(\frac{1}{2}+s)     \cr 
                            0             \cr 
                            0             \cr 
                           \frac{1}{2}-s        }
         \right)\ ,
      \label{wonefour}\\[8pt]
   {W_2}&=&\left(\matrix{ 0        \cr 
                         -1        \cr 
                          0        \cr 
                          0           }
         \right)\ .
         \label{Wtwofour}
\end{eqnarray}
The $X_i$ of Eqs.\ (\ref{xonefour}) and (\ref{xtwofour}) satisfy the same
commutation relations (\ref{algebra}) of the $L_1$ and $L_2$, namely,
\begin{equation}
\left[X_1,X_2\right]= X_2\ .
\label{algebraX}
\end{equation}

Equations (\ref{xonefour})--(\ref{Wtwofour}) are very instructive because
they exhibit the basic features due to the fact that the gauge group under
consideration is not semisimple. These features can be seen as follows.
Suppose a linear transformation $T$ is applied to $A_{\mu}$ of Eq.\
(\ref{reading}) so that the first two components of $TA_{\mu}$ are linear
combinations of the gauge fields $A_{\mu}^{(1)}$ and $A_{\mu}^{(2)}$ of
Eq.\ (\ref{usual}). Furthermore, after applying this $T$, $W_1$ and $W_2$
both take the form where the third and fourth components are zero. Let
$TX_2T^{-1}$ be expressed as
\begin{equation}
TX_2T^{-1}=\left({\matrix{Y_{11}&Y_{12}     \cr \noalign{\vskip3pt} 
                            0   &Y_{22}}}\right)\ ,
\label{Tchange}
\end{equation}
where the $Y$'s are $2\times 2$ matrices. While $Y_{21}$ is zero, the
question is: Can $Y_{12}$ be made zero or not? Consider any vector with
$v_4\mbox{$\,\ne\,$} v_1$
\begin{equation}
V_0=\left({\matrix{v_{1}\cr 
                   0   \cr 
                   0   \cr 
                   v_{4} }}\right)\ ,
\label{fourV}
\end{equation}
which is not a multiple of $W_1$. That
\begin{equation}
X_2V_0=\left({\matrix{0         \cr 
                      v_4-v_1   \cr 
                      0         \cr 
                      0 }}\right)
\label{fourVtf}
\end{equation}
means that all these $V_0$'s, which do not lie in the space spanned by
$W_1$ and $W_2$, give a $X_2V_0$ which is proportional to $W_2$. This
implies that
\begin{equation}
Y_{12}\mbox{$\,\ne\,$} 0\ .
\label{nonzeroY}
\end{equation}

That $Y_{12}$ is not zero has profound consequences. If the above
considerations are applied to a gauge group that is semisimple, the
resulting $Y_{12}$ can always be put to zero. Thus (\ref{nonzeroY}) is a
novel feature intimately related to the fact that the present gauge group
is not semisimple. Physically, that $Y_{12}$ is non-zero means that the two
additional gauge fields are not ``irrelevant'' and are coupled to the two
original gauge fields.

In order to see these new features more clearly, it is convenient to use
the following specific linear transform $T$:
\begin{equation}
\widetilde{A}_{\mu}^a=T^{ab} A_{\mu}^b
\label{Ttf}
\end{equation}
with
\begin{equation}
T=\left(\matrix{ 0        & 1         & 0      & 0           \cr
                 1        & 0         & 0      & -1          \cr
                 \frac{1}{2}-s  & 0         & 0      & \frac{1}{2}+s          \cr
                 0        & 0         & 1      & 0     }
    \right)\ .
\label{basistf}
\end{equation}
Note that this $T$ is of determinant one and hence invertible whatever be
the value of $s$. One finds that, in this $tilde$ basis, the infinitesimal
$\widetilde{A}_{\mu}$ transformation now becomes
\begin{equation}
\widetilde{A}^{\prime}_{\mu}=\widetilde{A}_{\mu}
+\alpha_i \widetilde{X}_i \widetilde{A}_{\mu}
+\partial_{\mu}\alpha_i \widetilde{W}_i
\label{Ahattf}
\end{equation}
with
\begin{eqnarray}
\widetilde{X}_1&=&\left(\matrix{ 1        & 0         & 0      & 0          \cr
                          0        & 0         & 0      & 0          \cr
                          0        & 0         & 0      & 0          \cr
                          0        & 0         & 0      & -1     }
         \right)\ ,
      \label{xonewt}\\[8pt]
\widetilde{X}_2&=&\left(\matrix{ 0        & -1        & 0      & 0          \cr
                          0        & 0         & 0      & 2          \cr
                          0        & 0         & 0      & -2s        \cr
                          0        & 0         & 0      & 0     }
         \right)\ ,
      \label{xtwowt}\\[8pt]
\widetilde{W}_1&=&\left(\matrix{0        \cr 
                         -1       \cr 
                         0        \cr 
                         0             }
         \right)\ ,
      \label{wonewt}\\[8pt]
\widetilde{W}_2&=&\left(\matrix{  -1      \cr 
                            0      \cr 
                            0      \cr 
                            0        }
         \right)\ .
      \label{wtwowt}
\end{eqnarray}
With the $T$ of Eq.\ (\ref{basistf}), the $Y_{12}$ of Eq.\ (\ref{Tchange})
is explicitly
\begin{equation}
Y_{12}=\left( {\matrix{0 & 0 \cr 
                       0 & 2}} \right)
\label{Yonetwo}
\end{equation}
which is not zero and cannot be brought to zero by any further change of
basis respecting (\ref{wonewt}) and (\ref{wtwowt}).

\section{Elementary Properties of Gauge Fields}\label{sec:3}

In the representation (\ref{rep2}) for $L_1$ and $L_2$, there is a
continuous parameter $s$. This is another feature not present for
finite-dimensional representations of semisimple groups.

In the transformations of the gauge fields $A_{\mu}$ and $\widetilde{A}_{\mu}$ as
given by (\ref{Afourtf}) and (\ref{Ahattf}) respectively, this parameter $s$
appears explicitly: in $W_1$ for $A_{\mu}$ and in $\widetilde{X}_2$ for
$\widetilde{A}_{\mu}$. Such appearances are undesirable because they imply that
both $A_{\mu}$ and $\widetilde{A}_{\mu}$ depend on the representation for the
doublet scalar field $\Phi(x)$ of Eq.\ (\ref{doublet}).

We therefore look for a further linear transform of the gauge field such
that this $s$ dependence appears in neither the new $X_i$ nor the new
$W_i$. For this purpose, consider first the case
\begin{equation}
s\mbox{$\,\ne\,$} 0\ .
\label{sneqzero}
\end{equation}
Under this assumption, define similar to Eq.\ (\ref{Ttf}),
\begin{equation}
\widehat{A}_{\mu}^a=\widetilde{T}^{ab} \widetilde{A}_{\mu}^b
\label{newTtf}
\end{equation}
with
\begin{equation}
\widetilde{T}=\left(\matrix{ 1        & 0         & 0      & 0           \cr
                      0        & 1         & 0      & 0           \cr
                      0        & 0         & 1/s    & 0           \cr
                      0        & 0         & 0      & 1     }
    \right)\ .
\label{newbasistf}
\end{equation}
The determinant of this diagonal $\widetilde{T}$ is $1/s$, which is well-defined
because of (\ref{sneqzero}). In the {\it{hat}} basis, the infinitesimal
transformation is
\begin{equation}
\widehat{A}^{\prime}_{\mu}=\widehat{A}_{\mu}
+\alpha_i \widehat{X}_i \widehat{A}_{\mu}
+\partial_{\mu}\alpha_i \widehat{W}_i
\label{newAhattf}
\end{equation}
with
\begin{eqnarray}
\widehat{X}_1=\widetilde{X}_1
        &=&\left(\matrix{ 1        & 0         & 0      & 0          \cr
                          0        & 0         & 0      & 0          \cr
                          0        & 0         & 0      & 0          \cr
                          0        & 0         & 0      & -1     }
         \right)\ ,
      \label{newxonewt}\\[8pt]
\widehat{X}_2
        &=&\left(\matrix{ 0        & -1        & 0      & 0          \cr
                          0        & 0         & 0      & 2          \cr
                          0        & 0         & 0      & -2         \cr
                          0        & 0         & 0      & 0     }
         \right)\ ,
      \label{newxtwowt}\\[8pt]
\widehat{W}_1=\widetilde{W}_1
         &=&\left(\matrix{0        \cr 
                         -1       \cr 
                         0        \cr 
                         0             }
         \right)\ ,
      \label{newwonewt}\\[8pt]
\widehat{W}_2=\widetilde{W}_2
        &=&\left(\matrix{  -1      \cr 
                            0      \cr 
                            0      \cr 
                            0        }
         \right)\ .
      \label{newwtwowt}
\end{eqnarray}
There is no dependence on $s$ in Eqs.\
(\ref{newxonewt})--(\ref{newwtwowt}), as desired. The $Y_{12}$ is still
given by (\ref{Yonetwo}).

The $\widehat{A}_{\mu}$, not $A_{\mu}$ or $\widetilde{A}_{\mu}$, is the desired
gauge field. Let the gauge transform (\ref{newAhattf}) be written out
component by component:
\begin{eqnarray}
\widehat{A}^{1\prime}_{\mu}&=&\widehat{A}_{\mu}^{1}
+\alpha_1 \widehat{A}_{\mu}^{1}
-\alpha_2 \widehat{A}_{\mu}^{2}
-\partial_{\mu}\alpha_2\ ,
\label{newAtfexplicit1} \\
\widehat{A}^{2\prime}_{\mu}&=&\widehat{A}_{\mu}^{2}
+2\alpha_2  \widehat{A}_{\mu}^{4}
-\partial_{\mu}\alpha_1\ ,
\label{newAtfexplicit2} \\
\widehat{A}^{3\prime}_{\mu}&=&\widehat{A}_{\mu}^{3}
-2\alpha_2  \widehat{A}_{\mu}^{4}\ ,
\label{newAtfexplicit3} \\
\widehat{A}^{4\prime}_{\mu}&=&\widehat{A}_{\mu}^{4}
-\alpha_1 \widehat{A}_{\mu}^{4}\ .
\label{newAtfexplicit4}
\end{eqnarray}
It is seen from (\ref{newAtfexplicit1}) and (\ref{newAtfexplicit2}) that
$\partial_{\mu}\alpha_2$ and $\partial_{\mu}\alpha_1$ appear respectively
in the gauge transforms of $\widehat{A}_{\mu}^{1}$ and
$\widehat{A}_{\mu}^{2}$, similar to those in the original paper of Yang and
Mills \cite{YM}. We therefore refer to these two components
$\widehat{A}_{\mu}^{1}$ and $\widehat{A}_{\mu}^{2}$ of the gauge fields as
\textit{Yang-Mills fields of the first kind}. In contrast, the derivatives
of $\alpha_1$ and $\alpha_2$ do not appear in the gauge transformations of
$\widehat{A}_{\mu}^{3}$ and $\widehat{A}_{\mu}^{4}$, as given by Eqs.\
(\ref{newAtfexplicit3}) and (\ref{newAtfexplicit4}). This is a new feature,
and we call these two components $\widehat{A}_{\mu}^{3}$ and
$\widehat{A}_{\mu}^{4}$ of the gauge fields \textit{Yang-Mills fields of
the second kind}.

In this present case of a non-semisimple gauge group, these two Yang-Mills
fields of the second kind are {\textit{not}} ``irrelevant.'' As seen from
Eq.\ (\ref{newAtfexplicit2}), $\widehat{A}_{\mu}^{4}$ appears on the
right-hand side, and thus plays a role in the gauge transform of the
Yang-Mills field $\widehat{A}_{\mu}^{2}$ of the first kind. This is the
direct consequence of the fact that the $Y_{12}$ of Eq.\ (\ref{Tchange}) is
not zero.

Aside from a possible linear transform among them, the Yang-Mills gauge
fields of the second kind are well defined. When the Yang-Mills fields of
the second kind are present, it is allowed to add arbitrary linear
combinations of these fields of the second kind to the Yang-Mills gauge
fields of the first kind.

It only remains to express the original $A_{\mu}$ in terms of these
Yang-Mills fields $\widehat{A}_{\mu}$:
\begin{equation}
A_{\mu}^a=R^{ab}\widehat{A}_{\mu}^{b}
\label{inversetf}
\end{equation}
with, because of (\ref{sneqzero}),
\begin{equation}
R=T^{-1}\widetilde{T}^{-1}=
\left(\matrix{   0        & s+\frac{1}{2}   & s      & 0           \cr
                 1        & 0         & 0      & 0           \cr
                 0        & 0         & 0      & 1          \cr
                 0        & s-\frac{1}{2}   & s      & 0     }
    \right)\ .
\label{inversebasistf}
\end{equation}
Of course, $A_{\mu}$ of (\ref{inversetf}) is to be used in the derivative
$D_{\mu}\Phi$ of Eq.\ (\ref{covder}), and depends on the value of $s$ of the
representation for $\Phi$, as expected.

Two comments are appropriate at this point. First, the only transform that
leaves $\widehat{X}_1,\widehat{X}_2,\widehat{W}_1$ and $\widehat{W}_2$
unchanged (sometimes referred as the stability group) is the identity.

Secondly, it is seen from Eq.\ (\ref{inversebasistf}) that the limit
$s\rightarrow 0$ is well defined for $R$ itself, and thus it is trivial to
remove the restriction (\ref{sneqzero}). In fact,
\begin{equation}
R\mid_{s\rightarrow 0}\mbox{}=
\left(\matrix{   0        & \frac{1}{2}     & 0      & 0           \cr
                 1        & 0         & 0      & 0           \cr
                 0        & 0         & 0      & 1          \cr
                 0        & -\frac{1}{2}    & 0      & 0     }
    \right)\ .
\label{inversebasistfszero}
\end{equation}
In this case of the representation with $s=0$, one Yang-Mills gauge field
of the second kind, $\widehat{A}_{\mu}^{3}$, decouples.

\section{The Diagonal $L_1$ Representations}\label{sec:4}  

In this section, we describe briefly the non-decomposable representations
of our algebra (\ref{algebra}) for which $L_1$ can be diagonalized. Those of
dimension $d=n+1$ are, up to a change of basis, of the form
\begin{eqnarray}
L_1&=&\left(\matrix{
          \frac{n}{2}+s & 0         &   0       & \ldots & 0          \cr
          0        &\frac{n}{2}-1+s &   0       & \ldots & 0          \cr
          0        &0          &\frac{n}{2}-2+s & \ldots & 0          \cr
          \vdots   &\vdots     &\vdots     & \ddots & \vdots     \cr
          0        & 0         &   0       & \ldots &-\frac{n}{2}+s     }
    \right)\ ,
    \nonumber\\[8pt]
L_2&=&\left(\matrix{0       & 1       & 0       & \ldots & 0      \cr
                    0       & 0       & 1       & \ldots & 0      \cr
                    0       & 0       & 0       & \ldots & 0      \cr
                    \vdots  & \vdots  & \vdots  & \ddots & \vdots \cr
                    0       & 0       & 0       & \ldots & 0 }
    \right)\ ,
\label{repnor}
\end{eqnarray}
where $s$ is an arbitrary complex parameter.

Let us briefly outline the arguments which can be used to prove this result.

Take any representation for which $L_1$ is diagonal. The eigenvalues can be
classified in sets associated with the following $s_i$'s:
\begin{equation}
\matrix{
s_1 &  , &  s_1+1 & , &  s_1+2 &  ,&  \ldots&  ,&  s_1+n_1 \cr 
s_2&   ,&  s_2+1&  ,  &  s_2+2 &  ,&  \ldots&  ,& 
s_2+n_2    \cr 
\vdots& &\vdots &     &\vdots  &   &\ddots  &   & \vdots    \cr 
s_i&  ,&  s_i+1&  ,&  s_i+2&  ,&  \ldots&  ,&  s_i+n_i
}\ ,
\label{ssets}
\end{equation}
where either $s_i-s_j$ for $i\mbox{$\,\ne\,$} j$ is not an integer or, if $s_i-s_j$ is
an integer, the corresponding sets are separated by at least 2 units.

Since, for any vector $V$ which is the eigenvector of $L_1$ with
eigenvalue $s$, we have by the basic commutator relations
\begin{equation}
L_1\left(L_2V\right)=\left(L_2L_1 +L_2\right)V=(s+1)\left(L_2V\right)
\ .
\label{schange}
\end{equation}
We see that two states can be connected by $L_2$ only in the case that they
belong to one of the above sets (\ref{ssets}). Hence, once transformed in
the block diagonal form corresponding to the above sets of eigenvalues, the
representation decomposes into these blocks. It is sufficient to study each
block in turn.

We call $s$ the corresponding lowest eigenvalue and $s+n$ its highest.
Suppose that the eigenvalue $s+k$ with $k=0,\ldots,n$ has multiplicity
$m_k\geq 1$.

Take any non-zero vector $V$ eigenstate of $L_1$ of eigenvalue $s+k$. The
successive action of $L_2$ on $V$ generates a set $\{S_V\}$ of non-zero
vectors
\begin{eqnarray}
&&L_1V=(s+k)V\ ,
\nonumber\\
&&\{S_V\}\equiv\left\{V,L_2V,(L_2)^2V,\ldots,
(L_2)^{\text{\textsf{L}}-1}V\right\}\ ,
\nonumber\\
&&(L_2)^{\text{\textsf{L}}}V=0
\ ,\quad 1\leq {\textsf{L}}\leq n-k\ .
\label{chain}
\end{eqnarray}
Let us call ${\textsf{L}}_V={\textsf{L}}$ the length of the chain built on
$V$.

We now consider the sequence $V_1,V_2,\ldots$ of vectors defined as follows:
\begin{itemize}
\item
Among all the vectors which are linear combinations of the eigenvectors of
$L_1$ with eigenvalue $s$, take any vector of minimal length
${\textsf{L}}_{V_1}$ and call it $V_1$.
\item
Among all the vectors which are linear combinations of the eigenvectors of
$L_1$ with eigenvalue $s$ but not along $V_1$, take any vector of minimal
length and call it $V_2$. Note that the length ${\textsf{L}}_{V_2}$ is
larger or equal to the length ${\textsf{L}}_{V_1}$.
\item
Among all the vectors which are linear combinations of the eigenvectors of
$L_1$ with eigenvalue $s$ but not situated in the subspace spanned by the
vectors $V_1,V_2$, take any vector of minimal length and call it $V_3$.
\item
Continue the process until the space of eigenvalue $s$ is exhausted, thus
defining successively $m_0$ vectors.
\item
Among all the vectors which are linear combinations of the eigenvectors of
$L_1$ with eigenvalue $s+1$ but not situated in the subspace spanned by the
vectors $\{L_2V_1,L_2V_2,\ldots,L_2V_{m_0}\}$, take any vector of minimal
length and call it $V_{m_0+1}$. If there is no such vector move to the
space of eigenvalue $s+2$ and repeat the process. Note that, this time, the
length
${\textsf{L}}_{V_{m_0+1}}$ can be smaller than the preceding lengths.
\item
Among all the vectors which are linear combinations of the eigenvectors of
$L_1$ with eigenvalue $s+1$ but not situated in the subspace spanned by the
vectors $\{L_2V_1,L_2V_2,\ldots,L_2V_{m_0},V_{m_0+1}\}$, select a vector of
minimal length and call it $V_{m_0+2}$. If there is no such vector move to
the space of eigenvalue $s+2$ and repeat the process excluding the subspace
$\{L_2^2V_1,L_2^2V_2,\ldots,L_2^2V_{m_0},L_2V_{m_0+1}\}$.
\item
Repeat the process successively for all the eigenvectors of eigenvalues
$s+1,s+2,s+3,\ldots$ until the complete space of all eigenvalues of $L_1$
is exhausted.
\item
Remark: This procedure provides a set of vectors
\begin{equation}
\left\{V_1,L_2V_1,L_2^2V_1,\ldots,V_2,L_2V_2,\ldots\right\}
\end{equation}
which are linearly independent. Indeed, if they were not, one would have a
combination of vectors of given eigenvalue $s+k$ (see (\ref{eigenequal})) of
$L_1$ equal to zero:
\begin{eqnarray}
&&\alpha_p
(L_2)^{a_p} V_p
+\alpha_q
(L_2)^{a_q} V_q
+\ldots
+\alpha_r
(L_2)^{a_r} V_r
+\ldots
+
(L_2)^{a_x} V_x
=0\ ,
\nonumber\\
&&\qquad {\rm{for}}\quad
p<q<\ldots<r<\ldots<x         \ .
\label{linearcomb}
\end{eqnarray}
Note that, if $V_r$ corresponds to an eigenvalue $s_r$ of $L_1$, the
integers $a_p,a_q,\ldots$ obey the relations
\begin{eqnarray}
&&s_p+a_p=s_q+ a_q =\ldots=s_r+a_r= \ldots=s_x+ a_x=s+k\ ,
  \nonumber\\
&&s_p\leq s_q\leq\ldots \leq s_r \leq \ldots\leq s_x\ ,
  \nonumber\\
&&a_r < {\textsf{L}}_{V_r}\ ,\quad{\rm{for\ all\ }} r\ .
\label{eigenequal}
\end{eqnarray}

By constructing the set $\{S_{V_x'}\}$ based on the vector
\begin{eqnarray}
V_x'&=&
\alpha_p
(L_2)^{a_p-a_x} V_p
+\alpha_q
(L_2)^{a_q-a_x} V_q
+\alpha_r
(L_2)^{a_r-a_x} V_r
+\ldots
+V_x\ ,
   \nonumber\\
L_2^{a_x}V'_x&=&0
\label{Vchange}
\end{eqnarray}
rather than $V_x$, one would construct the vector $V_x'$ of length $a_x$
smaller than the length of $V_x$ (see (\ref{eigenequal})) contrary to the
hypothesis.
\item
To any vector in the constructed series $V_r$, there corresponds the set
$\{S_{V_r}\}$, the basis of a non-decomposable representation of the group
of the form (\ref{repnor}) with a dimension $d={\textsf{L}}_{V_r}$ and with
a well-chosen
\begin{equation}
s=s_r+\frac{{\textsf{L}}_{V_r}-1}{2}  \ .
\end{equation}
\end{itemize}

This ends the proof of the decomposition of the representations for which
$L_1$ is diagonal.

We end this section with a word on the non-decomposable representations
where $L_1$ assumes a non-diagonal form, in fact a Jordan form. There are
many such representations. Some have a very elaborate structure. It is
sufficient for our later purpose to write the simplest example which is
3-dimensional and assumes the form
\begin{eqnarray}
L_1&=&\left(\matrix
  {\frac{2}{3}+s & 0                &  0          \cr 
     0           & -\frac{1}{3}+s   &  1          \cr 
     0           & 0                &-\frac{1}{3}+s }
    \right)\ ,
    \nonumber\\[8pt]
L_2&=&\left(\matrix{0    & 0    & 1      \cr 
                    0    & 0    & 0      \cr 
                    0    & 0    & 0  }
    \right)\ .
\label{repspe}
\end{eqnarray}

\section{Gauge Fields for Matter Fields Belonging to a\protect\\ diagonal
$L_1$ Representation}\label{sec:5}

In this section, we give the arguments showing that, for a matter field
belonging to a general diagonal $L_1$, non-decomposable, $d$-dimensional
representation, the gauge field structure is exactly the same as for a
matter field transforming with the two-dimensional representation
(\ref{rep2}). It consists of two gauge fields of the first kind and two
gauge fields of the second kind. There are $d^2-4$ other gauge fields which
are ``irrelevant'' as they decouple from the gauge fields of the first and
the second kinds.

Suppose that the matter field $\Phi(x)$ has $d$ scalar components and
transforms as in (\ref{Phitf}), (\ref{Vtf}) with the infinitesimal $L_1,L_2$
given by (\ref{repnor}). In the generalized derivatives there appears a set
of four $d\times d$ matrices $A_{\mu}$ transforming as (\ref{Atf}),
(\ref{Atfinf}). It is again convenient to associate the matrix $A_{\mu}$,
with components $A_{\mu}^{ij},\ i,j=1,\ldots,d$, with the $d^2$-dimensional
vector $A_{\mu}^a,\ a=1,\ldots,d^2$
\begin{equation}
A_{\mu}\Longrightarrow A_{\mu}^a
\label{matrixvector}
\end{equation}
by
\begin{equation}
           A_{\mu}^{d(i-1)+j}=A_{\mu}^{ij}\ .
\label{genbasis}
\end{equation}

For these $A_{\mu}^a$ fields, the gauge transformations are again of the
form (\ref{Afourtf}). The $X_i$ are $d^2\times d^2$ matrices generalizing
Eqs.\ (\ref{xonefour}) and (\ref{xtwofour}). The two $d^2$-dimensional
vectors
$W_1$ and $W_2$ are those associated with $-L_2$ and $-L_1$, respectively.
It is however somewhat easier to continue to work with the matrix
$A_{\mu}^{ij}$, using the initial $L_1$ and $L_2$ matrices (\ref{repnor})
and the commutator action (\ref{Atfinf}).

Let us introduce the following four $d\times d$ matrices, $M_m,m=-1,0,1$
and $P_0$, and thus the corresponding vectors. The non-zero elements of
these matrices are
\begin{eqnarray}
M_{-1}(j+1,j)=&{\displaystyle -\ \frac{(d-j)j}{2}} &,\quad j=1,\ldots,d-1\ ,
      \nonumber\\
M_{0}(j,j)=&{\displaystyle -\ \frac{d-2j+1}{2}} &,\quad j=1,\ldots,d\ ,
      \nonumber\\
M_{1}(j,j+1)=&1 &,\quad j=1,\ldots,d-1\ , 
      \nonumber\\
P_{0}(j,j)=&1 &,\quad j=1,\ldots,d\ .
\label{specialV}
\end{eqnarray}
These matrices obey commutation rules with $L_1$ and $L_2$ which govern the
homogeneous part of the transformation rules of the gauge vectors (see
(\ref{Atfinf})):
\begin{itemize}
\item
The three $M_m$ transform infinitesimally as a
3-dimensional representation of our algebra
\begin{eqnarray}
\left[L_1,M_m\right]&=&mM_m\ ,
    \nonumber\\
\left[L_2,M_1\right]&=&0\ ,
    \nonumber\\
\left[L_2,M_m\right]&=&M_{m+1}\ ,\quad m=-1,0\ ,
\label{genAtfspin1}
\end{eqnarray}
\item
while the last one, $P_0$, transforms as the 1-dimensional representation
\begin{eqnarray}
\left[L_1,P_0\right]&=&0\ ,
    \nonumber\\
\left[L_2,P_0\right]&=&0\ .
\label{genAtfspin0}
\end{eqnarray}
\end{itemize}
Finally, we have the correspondence
\begin{eqnarray}
L_1=&-M_0+sP_0&\Longrightarrow  -W_2\ ,
    \nonumber\\
L_2=&M_1      &\Longrightarrow  -W_1\ .
\label{correspond}
\end{eqnarray}

The gauge vectors which lie in the direction of these four vectors are the
four ``relevant''  gauge fields. They contain the gauge fields of the first
kind in the directions of $W_1$ and $W_2$ and the two gauge fields of the
second kind. The $d^2-4$ remaining vectors can be classified by using the
reduction of a general representation as described in Sec.\ \ref{sec:4}
starting the procedure with the vector $V_1$ of lowest value of $s=
-(d-1)/2$ corresponding to the matrix with only one non-zero element,
namely, $V_1(d,1)=1$.

Using the notation $[d]$ for a representation of dimension $d$, it is easy
to see that the resulting decomposition of the action on the $d^2$ gauge
fields (\ref{genbasis}) is as follows:
\begin{equation}
[d^2]=\sum_{\oplus i=0}^{d-1}[2i+1]=[1]\oplus[3]
\oplus \sum_{\oplus i=2}^{d-1}[2i+1]\ ,
\label{decomp}
\end{equation}
where the part $[4]=[1]\oplus[3]$ is equivalent to (\ref{specialV}) and, in
the basis just referred to, acts on the above 4-dimensional space which
contains the subspace spanned by vectors $W_1$ and $W_2$. The other
$(2i+1)$-dimensional representations (with $2\leq i\leq d-1$) are the
non-decomposable ones of the diagonal case ((\ref{repnor}) with their own
$s_i=0$).

In a way completely analogous to the argument outlined around
(\ref{Tchange}), we apply a transformation $T$ in such a way as to bring
the linear combinations of the four vectors (\ref{specialV}) in positions
$1, 2, 3, 4$ and to put the $d^2-4$ other vectors obtained in the procedure
in the remaining positions $5,\ldots,d^2$. Then, both the $d^2\times d^2$
matrices $X_1$ and $X_2$ take the form
\begin{equation}
TX_iT^{-1}=\left({\matrix{Z^i_{11}&0    \cr\noalign{\vskip3pt}
                            0   &Z^i_{22}}}\right)\ ,
\label{genTchange}
\end{equation}
where $Z^i_{11}$ is a $4\times 4$ matrix analogous to (\ref{Tchange}) and
$Z^i_{22}$ a $(d^2-4)\times (d^2-4)$ matrix. The non-diagonal blocks are
zero.

Hence we conclude that all matter fields transforming with diagonal
representations belong to the same theory with, in the general case,
exactly the \textit{same four} vector fields (\ref{specialV}). Again there
are two Yang-Mills gauge fields of the first kind and two Yang-Mills gauge
fields of the second kind. The gauge fields corresponding to the $d^2-4$
remaining vectors are ``irrelevant'' as seen from (\ref{genTchange}).

To conclude this section let us write explicitly the matrix $A_{\mu}$ which
has to be used in the covariant derivative (\ref{covder}) of the
$d$-dimensional field
\begin{eqnarray}
A_{\mu}&=&\widehat{A}_{\mu}^{1}M_1-\widehat{A}_{\mu}^{2}M_0
           -2\widehat{A}_{\mu}^{4}M_1
        +s(\widehat{A}_{\mu}^{2}+\widehat{A}_{\mu}^{3})P_0
    \nonumber\\
        &=&\widehat{A}_{\mu}^{2}L_1+\widehat{A}_{\mu}^{1}L_2
           +s\widehat{A}_{\mu}^{3}P_0
           -2\widehat{A}_{\mu}^{4}M_1  \ .
\label{AtotalX}
\end{eqnarray}
Again we see that for $s=0$, the field $\widehat{A}_{\mu}^{3}$ decouples.

{}From Eq.\ (\ref{AtotalX}), we see the correspondence between the two
Yang-Mills gauge fields usually called $A_{\mu}^{(1)}$ and $A_{\mu}^{(2)}$
(see
(\ref{usual})) and our {\it{hat}} fields $\widehat{A}_{\mu}^{1}$ and
$\widehat{A}_{\mu}^{2}$:
\begin{eqnarray}
A_{\mu}^{(1)}&=&\widehat{A}_{\mu}^{2}\ ,
   \nonumber\\
A_{\mu}^{(2)}&=&\widehat{A}_{\mu}^{1} \ .
\label{hatnonhat}
\end{eqnarray}

If an analogous study is performed using, for the scalar fields, another
non-decomposable representation with a non-diagonal $L_1$, the picture
changes drastically. We have analyzed in full detail what happens for a few
of these representations and in particular for the representation
(\ref{repspe}). In the later case there are, apart from the two Yang-Mills
fields of the first kind, in general seven Yang-Mills fields of the second
kind. This new system of altogether nine Yang-Mills fields does not contain
the set of the four Yang-Mills fields relevant to the representations where
$L_1$ is diagonal.

\section{The Gauge Field Lagrangian}\label{sec:6}

We can write easily a gauge invariant Lagrangian which
is, up to a factor,
\begin{equation}
{\cal{L}}^{\uppercase\expandafter{\romannumeral1}}
     ={\rm{trace\ }}\left(F^{\mu\nu}F_{\mu\nu}\right)
\label{usualL}
\end{equation}
familiar for simple groups but also
\begin{equation}
{\cal{L}}^{\uppercase\expandafter{\romannumeral2}}
     =\left({\rm{trace\ }}F^{\mu\nu}\right)
                   \left({\rm{trace\ }}F_{\mu\nu}\right) \ .
\label{usualLprime}
\end{equation}

These results are obvious if the curvatures are
defined, as usual, by
\begin{equation}
F_{\mu\nu}=\left[D_{\mu},D_{\nu}\right]\ ,
\label{curvs}
\end{equation}
which, in view of (\ref{covdertf}), transform as
\begin{eqnarray}
\left(F_{\mu\nu}\right)^{\prime}&=&VF_{\mu\nu}V^{-1}
      \nonumber\\
     &= & F_{\mu\nu} +\alpha_i\left[X_i,F_{\mu\nu}\right]
     \quad {\rm{infinitesimally}}\ .
\label{curvattf}
\end{eqnarray}
This means that the curvatures are invariant under the non-homogeneous part
of the transformation (\ref{Ahattf}) induced by the $W$'s and transform
covariantly under the homogenous part induced by $V$. The invariance of the
Lagrangians (\ref{usualL}) and (\ref{usualLprime}) follow. It should be
noted that these Lagrangians are invariant not only under our gauge group
but more generally under transformations with any matrix $V$, namely the
group GL(2,$R$) or even GL(2,$C$), which contain our group as a subgroup.

Since the $\widehat A_{\mu}^a,\ a=1,\ldots,4$ gauge fields are the basic
fields of the first and of the second kind for all the diagonal
representations, we focus our attention on them. Recall that their
infinitesimal transformation properties are summarized in (\ref{newAhattf})
with $X_i$ and $W_i$ given by (\ref{newxonewt})--(\ref{newwtwowt}).

In view of the new feature related to the presence of gauge fields of the
second kind, we did not want to be prejudiced by the familiar result and we
have decided to start with a minimal set of general conditions.
\begin{enumerate}

\item
The Lagrangian should be Lorentz invariant.

\item Terms of a kinetic energy type for the vector fields should appear in
the Lagrangian, i.e., a sum of terms quadratic in the space-time derivatives
of the fields
\begin{equation}
\beta_1(a,b)\, \left(\partial_{\mu}\widehat{A}_{\nu}^a\right)
                    \left( \partial^{\mu}\widehat{A}^{b\nu}\right)
        +\beta_2(a,b)\, \left(\partial_{\nu}\widehat{A}_{\mu}^a\right)
                     \left(\partial^{\mu}\widehat{A}^{b\nu}\right)
\label{type1}
\end{equation}
for a suitable set of values of the constants $\beta_i(a,b)$.

\item The Lagrangian should be invariant under the gauge transformations.

\item
Terms which are total divergences can be eliminated.

\end{enumerate}

As a result of condition 3 and because of the existence of the
inhomogeneous part in the transformation, two types of terms should be
added to (\ref{type1}):

\begin{itemize}
\item
terms quadratic in
the fields and at the same time linear in the space-time derivatives
\begin{equation}
\gamma(a,b,c)\, \left(\partial_{\nu}\widehat{A}^{a}_{\mu}\right)
                     \widehat{A}^{b\nu}\widehat{A}^{c\mu}\ ;
\label{type2}
\end{equation}

\item terms quartic in the vector fields
\begin{equation}
\delta(a,b,c,d)\, \widehat{A}^{a}_{\mu}\widehat{A}^{b\mu}
                     \widehat{A}^{c}_{\nu}\widehat{A}^{d\nu}
\ .
\label{type3}
\end{equation}

\end{itemize}

The coefficients $\beta_1(a,b)$, $\beta_2(a,b)$, $\gamma(a,b,c)$ and
$\delta(a,b,c,d)$ obviously enjoy the symmetries
\begin{eqnarray}
\beta_1(a,b)&=&\beta_1(b,a)\ ,
     \nonumber\\
\beta_2(a,b)&=&\beta_2(b,a)\ ,
     \nonumber\\
\delta(a,b,c,d)&=&\delta(b,a,c,d)\ ,
     \nonumber\\
\delta(a,b,c,d)&=&\delta(a,b,d,c)\ ,
     \nonumber\\
\delta(a,b,c,d)&=&\delta(c,d,a,b)\ .
\label{symlag}
\end{eqnarray}
Taking the most general linear combination of terms of the form
(\ref{type1}), (\ref{type2}) and (\ref{type3}), after lengthy computations,
we have shown that we recover, in general, a linear combination of the two
obvious Lagrangians (\ref{usualL})
and (\ref{usualLprime}) and nothing more.

More precisely:
\begin{itemize}
\item
First as a consequence of the space dependence of the parameters, i.e. to
the presence of the inhomogeneous terms in the transformation, we have
obtained the expected result: the Lagrangian can be written in terms of the
covariant derivatives $\widehat{F}^{\mu\nu a}$ only. The Lagrangian then
takes the form
\begin{equation}
\beta_{ab} \widehat{F}^{\mu\nu a}\widehat{F}_{\mu\nu}^{b}\ .
\label{Laggene}
\end{equation}

\item
Global symmetry then remains to be imposed. This leads to the final
restrictions
\begin{eqnarray}
&&\beta_{11}=\beta_{12}=\beta_{13}=\beta_{24}=\beta_{34}=\beta_{44}=0\ ,
      \nonumber\\
&&\beta_{14}=2(\beta_{22}-\beta_{33})\ ,
      \nonumber\\
&&\beta_{23}=\beta_{33}  \ .
\label{Lagcoef}
\end{eqnarray}
We see that there are two free parameters $g_1=\beta_{22}$ and
$g_2=\beta_{33}$.

\end{itemize}

The final form of the most general invariant Lagrangian in the {\it{hat}}
basis is then
\begin{eqnarray}
{\cal{L}}&=& g_1
\left( \widehat{F}^{(2)\mu\nu}\widehat{F}_{\mu\nu}^{(2)}
    +4\widehat{F}^{(1)\mu\nu}\widehat{F}_{\mu\nu}^{(4)}\right)
      \nonumber\\
&&\mbox{}+g_2\left(\widehat{F}^{(3)\mu\nu}\widehat{F}_{\mu\nu}^{(3)}
     +2\widehat{F}^{(2)\mu\nu}\widehat{F}_{\mu\nu}^{(3)}
     -4\widehat{F}^{(1)\mu\nu}\widehat{F}_{\mu\nu}^{(4)}\right)\ .
\label{Laggenefinal}
\end{eqnarray}

Going back from the {\it{hat}} basis to the initial basis for the
$A_{\mu}$'s by the $R$ transform of (\ref{inversetf}), we recover for the
particular values
\begin{equation}
g_1=\frac{5}{2}\ ,\quad g_2=2
\label{gl2g}
\end{equation}
the usual Lagrangian (\ref{usualL}) as a particular case, while for
\begin{equation}
g_1= g_2=4
\label{gl2gprime}
\end{equation}
we recover the second Lagrangian (\ref{usualLprime}).

We defer to a later work the study of the possible physical consequences of
the fact that, by choosing suitably $g_1$ and $g_2$, certain gauge fields
could have no kinetic energy (for example if $g_2=0$) and hence can be
eliminated from the equation of motion though they appear explicitly in the
Lagrangian.

\section{Conclusions and Discussion}\label{sec:7}

As shown almost fifty years ago by Yang and Mills [1] when they discovered
and introduced gauge theories, the number of gauge fields can always be
chosen to be equal to the number of the generators of the gauge group.  This
result is valid for any group.

Moreover, Yang and Mills already discussed in their original article the
possibility of additional gauge fields. They remarked that, for semisimple
groups and semisimple groups multiplied by U(1)'s, these additional gauge
fields can be removed from the theory because they are irrelevant. A way of
understanding their idea is to note that, for these groups, the gauge
transformations of the additional fields completely decouple from the
original Yang-Mills fields. In other words, Yang-Mills fields transform
among themselves and the additional gauge fields among themselves
separately.

In this paper, we address the issue of these additional gauge fields for 
non-semisimple groups.  It is found that, for this case as distinct from 
that of the semisimple groups possibly multiplied by U(1)'s, there can be
additional gauge fields that are not irrelevant.  That is, these 
additional gauge fields appear in the gauge transform of the original 
Yang-Mills fields.  In such cases, we refer, by definition, to the 
original Yang-Mills fields as Yang-Mills gauge fields of the first kind, 
and to the additional Yang-Mills fields as Yang-Mills gauge fields of the
second kind.  Yang-Mills gauge fields of the second kind are well defined,
but it is permitted to alter Yang-Mills gauge fields of the first kind by
adding  to them arbitrary linear combinations of Yang-Mills gauge fields of
the  second kind.

The case of the simplest non-semisimple group, where there are only two
generators $L_1$ and $L_2$ which satisfy  Eq.~(\ref{algebra}), is worked out
in detail.  In this case we have studied and determined explicitly what
happens  when the matter field transforms as a representation of that group
for which $L_1$ can be diagonalized.  Apart from the two Yang-Mills gauge
fields of the first kind corresponding to the two generators, the allowed
additional gauge fields separate into two Yang-Mills gauge fields of the
second kind with the remaining gauge fields being irrelevant. No change of
basis allows the decoupling of the gauge fields of the second kind and
hence the elimination of these additional gauge fields.

For matter fields belonging to representations of our group where $L_1$
cannot be diagonalized, the situation is much more complicated.  For only
one of these cases are the results reported briefly in this article.

We have shown that the Lagrangian for the gauge fields is not unique since
the most general gauge invariant Lagrangian depends on two arbitrary
parameters.  For certain values of the parameters, some gauge fields may
have no kinetic energy and appear in the theory as non-propagating fields
of spin 1. 

A general argument can be given as follows to indicate that the behavior
we have discovered for the above simplest group is generic.  For
semisimple groups, the reducible representations are all fully reducible
(or decomposable).  This is not true for non-semisimple groups.  Since the
gauge fields belong to the direct product of the representation of the
matter field with its inverse transposed, this product is always reducible
as it contains the adjoined representation to which the Yang-Mills gauge
fields of the first kind belong.  But since for non-semisimple groups this
product is in general reducible but not fully reducible, there will be
Yang-Mills gauge fields of the second kind connected to the Yang-Mills
fields of the first kind, in a way analogous, with appropriate changes, to
(\ref{Tchange}) and (\ref{nonzeroY}).

\vskip .7truecm
\noindent{\large\bf{Acknowledgment}}
\vskip 0.2truecm
\noindent
A large part of this work was carried out while the authors were visiting
CERN. We are very grateful to the Theory Division for its hospitality.

The work of one of the authors (J.N.) was supported in part by 
the Belgian Fonds National de la Recherche Scientifique, while that 
of the other (T.T.W.) was supported in part by the United States 
Department of Energy under Grant No.\ DE-FG02-84ER40158.\vadjust{\kern-2pt}

\end{document}